\documentclass[aip, amsmath,amssymb, graphicx, floatfix]{revtex4-1}

\usepackage[utf8]{inputenc}
\usepackage[T1]{fontenc}
\usepackage[english]{babel}
\usepackage{natbib}
\usepackage{hyperref}
\usepackage{array}
\usepackage{makecell}
\usepackage{multirow}
\usepackage{url}
\usepackage{etoolbox}
\usepackage{lineno}
\usepackage{nameref}
\usepackage{xr}
\usepackage{graphicx}
\usepackage{mhchem}

\draft 

\begin{document}


\title[]
  {
 Single acquisition reconstruction of nonlinear susceptibility and Raman tensors at the diffraction limit} 



\author{Volodymyr Multian}
\affiliation{Department of Quantum Matter Physics, University of Geneva, 24 Quai Ernest Ansermet, Geneva CH-1211, Switzerland}
\affiliation{Nonlinear BioImaging Lab, Department of Applied Physics, University of Geneva, 24 Quai Ernest Ansermet, Geneva CH-1211, Switzerland}
\affiliation{Photon Processes Department, Institute of Physics of NAS of Ukraine. Prospect Nauky, 46, Kyiv-03028, Ukraine.}

\author{Luigi Bonacina}
\affiliation{Nonlinear BioImaging Lab, Department of Applied Physics, University of Geneva, 24 Quai Ernest Ansermet, Geneva CH-1211, Switzerland}

\author{Jérémie Teyssier}
\affiliation{Department of Quantum Matter Physics, University of Geneva, 24 Quai Ernest Ansermet, Geneva CH-1211, Switzerland}
\email{jeremie.teyssier@unige.ch}


\date{\today}

\begin{abstract}

Quantitative extraction of nonlinear optical tensors at the microscale typically requires sequential polarization scans, mechanical rotations, and careful correction of high-NA polarization distortions. We introduce a single-acquisition tensor metrology approach that simultaneously acquires Raman and second-harmonic generation (SHG) signals and enables rapid reconstruction of both Raman and $\chi^{(2)}$ tensors from a single diffraction-limited volume.

The central innovation is the use of a tightly focused Bessel–Gauss (BG) beam whose intrinsic spatiotemporal structure encodes multiple excitation geometries—normal and finite incidence—within a single measurement. The resulting Fourier-plane SHG pattern contains rich polarization and angular information, eliminating mechanical rotation and reducing acquisition time.

Combined with a polarization-preserving dichroic module and GPU-accelerated ray-tracing global fitting, the method exploits the full two-dimensional SHG emission pattern together with polarization-resolved Raman data. We demonstrate quantitative tensor extraction in LiNbO$_3$ and correlate symmetry breaking across the ferroelectric transition of KH$_2$PO$_4$. This approach establishes rapid, calibration-robust optical tensor tomography for micro-scale materials.

\end{abstract}


\maketitle 


\section{Introduction}
Raman spectroscopy and Second Harmonic Generation (SHG) are powerful, complementary techniques that offer unique insights into the properties of materials. Each method provides distinct information, which, when combined, allows for a comprehensive understanding of a material's structural and electronic characteristics. Both are non-destructive, making them especially valuable for monitoring processes without altering the sample. This synergy is particularly beneficial for obtaining a comprehensive view of complex materials.

Polarization-resolved measurements significantly improve the specificity and sensitivity of both techniques. In Raman spectroscopy, polarization analysis enables the identification of vibrational mode symmetries\cite{loudon2001raman}. In SHG, polarization plays a crucial role in determining symmetry properties such as stacking order in two-dimensional materials or heterostructures, as well as crystal orientation\cite{heinz_study_1985,tom_second-harmonic_1983,yamada_anisotropy_1993}. When fully implemented, polarimetric SHG and Raman measurements allow extraction of the complete second-order susceptibility tensor and the full Raman tensors, respectively.

The simultaneous measurement of polarization-dependent Raman and SHG signals is applicable across diverse fields, including materials science \cite{cantarero2015raman, yokota2012optical}, biology\cite{chandra2024unveiling, bueno2016second}, chemistry\cite{fan2020review, geiger2009second}, and nanotechnology\cite{kumar2012raman, bonacina2020harmonic}. For instance, in biological tissues, Raman spectroscopy can provide biochemical information\cite{chandra2024unveiling}, while SHG can image structural proteins like collagen and myosin \cite{aghigh_second_2023}. This dual capability enhances the ability to study complex biological structures and dynamic processes.

In this work, we introduce a new experimental setup that combines Raman and SHG microscopy. We will present the specificity of both Raman and SHG sub-systems and the technical challenges for coupling these two techniques while preserving full polarization control for both excitation lasers and collected signals. A completely new approach is proposed to reliably and rapidly extract the nonlinear and Raman optical tensors. We will detail the experimental aspects of this implementation as well as data analysis, which relies on a polarization-resolved ray-tracing model for data fitting.

\section{Results and discussion}
\subsection{Nonlinear optical tensor extraction}\label{sec:RA-SHG}
\subsubsection{Experimental approach}

In the electrical dipole approximation, the nonlinear polarization $\vec{P}$, involved in SHG process, can be written as \cite{boyd_nonlinear_2008}:

\begin{equation}
    P_{i} = \varepsilon_0 \chi_{ijk}^{(2)}E_j E_k,
\label{eq:P}\end{equation} where $E_{i,j}$ refers to the electrical field of pump beam and $\chi_{ijk}^{(2)}$, the tensor of second order nonlinear susceptibility. Here, we use Einstein's summation formalism, which assumes the summation over repeated indices. Equation \ref{eq:P} describes nonlinear polarization P induced in the material by the plain wave $\vec E^{\omega}$, which is defined in 3D space of the laboratory frame. Corresponding electric field at $2\omega$ is proportional to $\vec P$ with coefficients that are defined by phase matching conditions, length of propagation path, coherence length, and direction of observation.. The tensor $\chi_{ijk}^{(2)}$ is a rank 3 tensor ($3\times3\times3$) which contains 27 complex elements. For a given compound, the structural point group defines identical and zero elements in the tensor. Higher symmetry implies a lower number of independent tensor elements.

\begin{figure*}[!ht]
    \centering
    \includegraphics[width=1\linewidth]{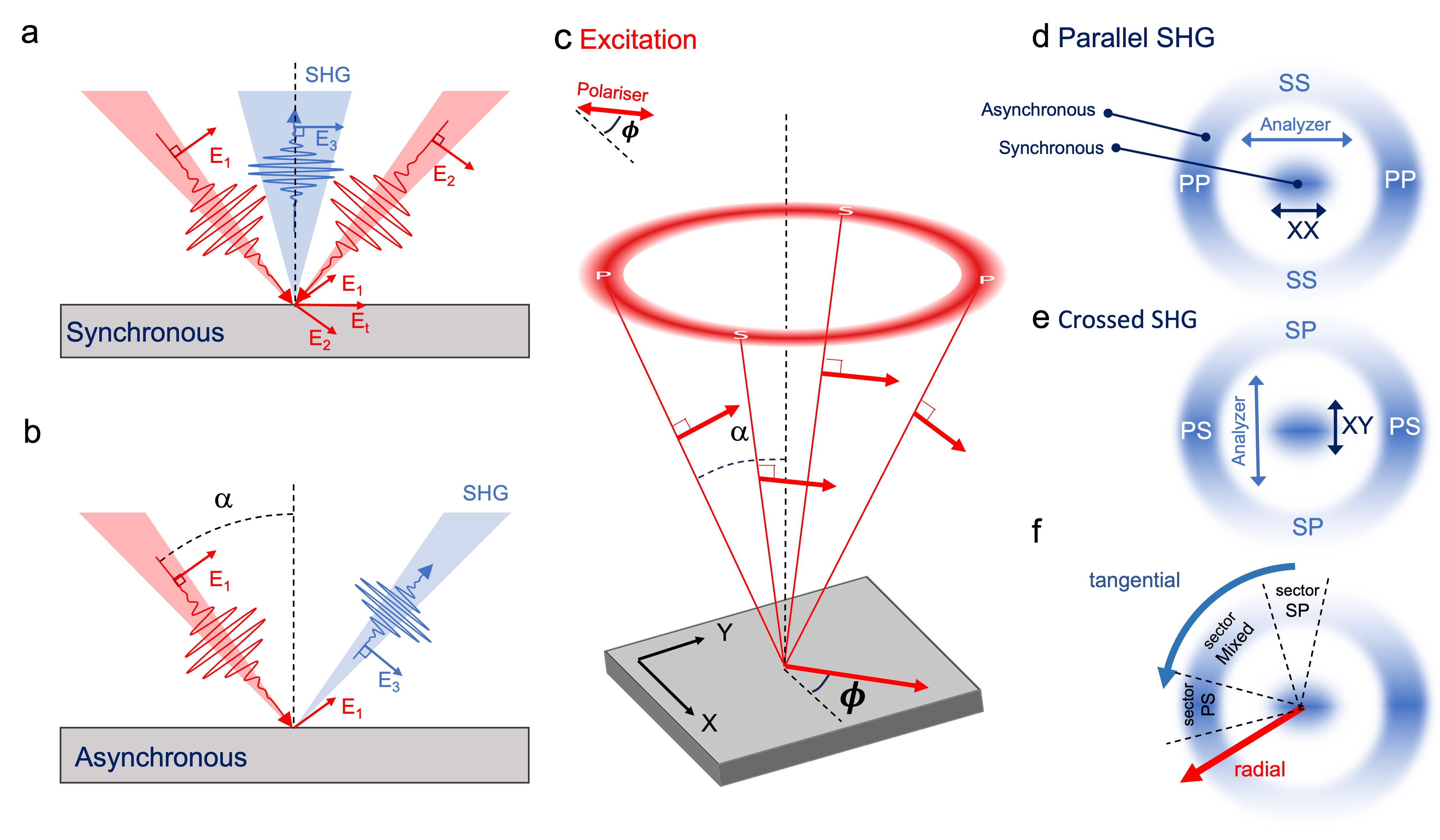}
    \caption{a) Polarization state at the focal point of BG beam excitation (in red) under a) synchronous (non-collinear) and b) asynchronous (collinear) conditions. c) Azimuthal dependence of the polarization state for the excitation BG beam polarized at an angle $\phi$ relative to the sample x-axis. d-e) Corresponding patterns and polarization configurations for the  (SHG) signal. The central part corresponds to the synchronous state probing normal incidence configuration XX, XY (S and P states are equivalent), while the ring is associated with the fixed incidence asynchronous polarization configuration. f) Definition of the different sectors and angles used for data analysis.}
    \label{fig:MRA_NHG}
\end{figure*}

 Rotational Anisotropy Second Harmonic Generation (RA-SHG) is a well-established experimental method to determine the quadratic nonlinear susceptibility tensor, $\chi^{(2)}$. Originally developed to study crystallographic and electronic symmetries in bulk materials, RA-SHG offers a rapid approach to characterize the SHG polarization dependence \cite{yamada_anisotropy_1993, heinz_study_1985, tom_second-harmonic_1983}. Specifically, a laser beam with either $P$ (parallel) or $S$ (perpendicular) polarization illuminates a sample at a fixed angle of incidence. SHG intensity is acquired while rotating the sample around the optical axis and varying the excitation/detection polarization settings ($P/P$, $P/S$, $S/P$, $S/S$). The polarization dependence of the resulting signal is fitted to a response calculated from the nonlinear susceptibility tensor and the specific experimental geometry. The accurate identification of the $\chi^{(2)}$ tensor elements depends on knowledge of the crystal’s symmetry (point group) and its orientation. Higher symmetry limits both the number of independent tensor components and the number of sample orientations required for a reliable fitting. Conversely, when the nonlinear tensor is known, it can be used to infer the crystal’s orientation.

In a typical RA-SHG setup, the sample is excited by a linearly polarized ($P$ or $S$) pump beam at an angle $\alpha$ to the surface normal, and the emitted SH signal is analyzed with the polarizer along the $S$ or $P$ orientations. A complete analysis requires rotating the sample around its normal axis, producing four polarization-resolved curves based on the input/output configuration: $P/P$, $P/S$, $S/P$, $S/S$. By fitting these curves, the $\chi^{(2)}$ tensor can be reconstructed. This method is effective for parallel or weakly focused laser beams but is limited in spatial resolution or sample size. For micron-scale resolution, RA-SHG must be adapted to microscope geometries \cite{rendon-barraza_crystalline_2019, schmidt_multi-order_2016, aghigh_second_2023, butet_three-dimensional_2010, schwung_nonlinear_2014}, which poses challenges such as precise sample rotation at the microscale. A common solution is rotating the laser beam's polarization using retardation optics \cite{lien_precise_2013}. Tight focusing required in high-resolution optical measurements involves objectives with high numerical aperture that introduce polarization distortions \cite{chen_calibration_2019, chen_chapter_2012}, causing mismatches between the polarization states at the focal point and at the back-focal aperture. Moreover, pulsed laser excitation needs precise wavefront control. Ultrashort laser pulses passing through different glass thicknesses experience wavefront distortion, leading to variations in pulse arrival times at the focused beam waist, resulting in polarization distortions \cite{shatrovoy_second-harmonic_2016}. These distortions can cause misinterpretation of polarization dependencies when using a single sensor. Minimizing time delays with adaptive optics \cite{salter_adaptive_2019} and addressing refraction-induced polarization distortions with tailored opto-mechanical solutions \cite{torchinsky_low_2014, lu_fast_2019, harter_high-speed_2015, thibon_resolution_2017, rao_conceptual_2024, arlt_efficiency_1999} are necessary. Even when the polarization state and the spatial phase profile of the incoming light are pre-compensated, the approach is limited to the excitation path but cannot correct for light emitted by the sample.

In this work, we introduce a new Microscopic Rotational Anisotropy Second Harmonic Generation (MRA-SHG) method that overcomes many of these challenges by exploiting intrinsic wavefront distortions, imperfect spatial and temporal focusing in tightly focused beams, and self-phase-matching effects \cite{netz_measurement_2000, caron_phase_1998, rao_conceptual_2024, arlt_efficiency_1999} to extract maximum information about 3D nonlinear optical response of the material. We propose leveraging the inherent imperfections in spatial and temporal focusing found in real optical systems, rather than compensating for them. This is achieved by maximizing the information extracted from a single-point measurement through the use of a Bessel-Gauss beam \cite{GORI1987491, mcgloin_bessel_2005, shatrovoy_second-harmonic_2016, caron_phase_1998}, in place of the conventional Gaussian laser beam profile \cite{thibon_resolution_2017,mcgloin_bessel_2005, arlt_efficiency_1999, rao_conceptual_2024}. The benefit of such a configuration is detailed in Figure~\ref{fig:MRA_NHG}. When a sample is excited by a BG-shaped beam, it emits a signal with a specific 3D distribution of intensity defined by the geometry of excitation, crystallographic orientation and tensor of response. This signal can be captured on a 2D imaging sensor placed in the Fourier plane of the objective. When photons from opposite sides of the optical axis coincide in time and space at the focal plane, interference occurs, resulting in an in-plane polarization component ($\vec E_t = \vec E_1 + \vec E_2$ in Figure~\ref{fig:MRA_NHG}a) and the emission of a harmonic signal normal to the surface. This emission is detected as a sharp peak in the center of the Fourier pattern (spatiotemporal overlap spot in Figure~\ref{fig:MRA_NHG}d,e). This portion of the image enables recording the polar map at normal incidence.

On the other hand, photons reaching the sample with unbalanced time-delay or having no spatial overlap with photons from opposite side of the optical axis, contribute to an out-of-plane excitation geometry (Figure~\ref{fig:MRA_NHG}b) and provide a measurement at a fixed angle of incidence ($\alpha$ in Figure~\ref{fig:MRA_NHG}b-c). This emission, with no spatiotemporal overlap, is collected as a ring on the 2D detector as shown in Figure~\ref{fig:MRA_NHG}d,e. To our best knowledge, the simultaneous acquisition of SHG upon both in- and out-of-plane excitation at a given polarization angle $\phi$ has never been exploited so far. This approach does not require any mechanical manipulations with the sample, reducing the measurement time for single polarization configuration ultimately to the integration time of the CCD camera. Total measurement time depends only on the set of incoming polarizations and orientations of the analyzer that can be minimized to a set of the most informative one analogically to the optical polarimetry\cite{del2022calibration}. The normal incidence dataset is obtained by integrating the central region of the Fourier plane image for each polarization angle $\phi$ (Figure~\ref{fig:MRA_NHG}d,e). It should be noted that, in principle, the described model could be applied for low NA systems, however it reduces the precision of out of plane components determination due to the smaller angles of incidence and worth decoupling of synchronous and asynchronous parts of the image in the Fourier plane.

In addition to providing a comprehensive analysis of high-symmetry point groups in the $PP$, $PS$, $SP$, and $SS$ maps, where only a limited number of tensor elements are involved, the use of ray tracing and evolutionary algorithms (see Section~\ref{subsec:data_analysis}) allows the full information contained in the 2D pattern to be exploited, rather than relying on averaged signals from the $PP$, $PS$, $SP$, and $SS$ sectors, as shown in Figure~\ref{fig:MRA_NHG}f. Indeed, the radial component contains the dependence of the signal to the angle of incidence $\alpha$ (red arrow in Figure~\ref{fig:MRA_NHG}f) and the tangential component provides a continuous variation from one high symmetry point to another. It can be computed and fitted integrally (blue arrow in Figure~\ref{fig:MRA_NHG}f), capitalizing on the information contained in mixed polarization states. The approach to collect and analyze this complex 2D pattern is detailed in the following sections.

\subsection{Experimental setup}\label{sec:setup}

\begin{figure}[!h]
    \centering    
    \includegraphics[width=0.8\linewidth]{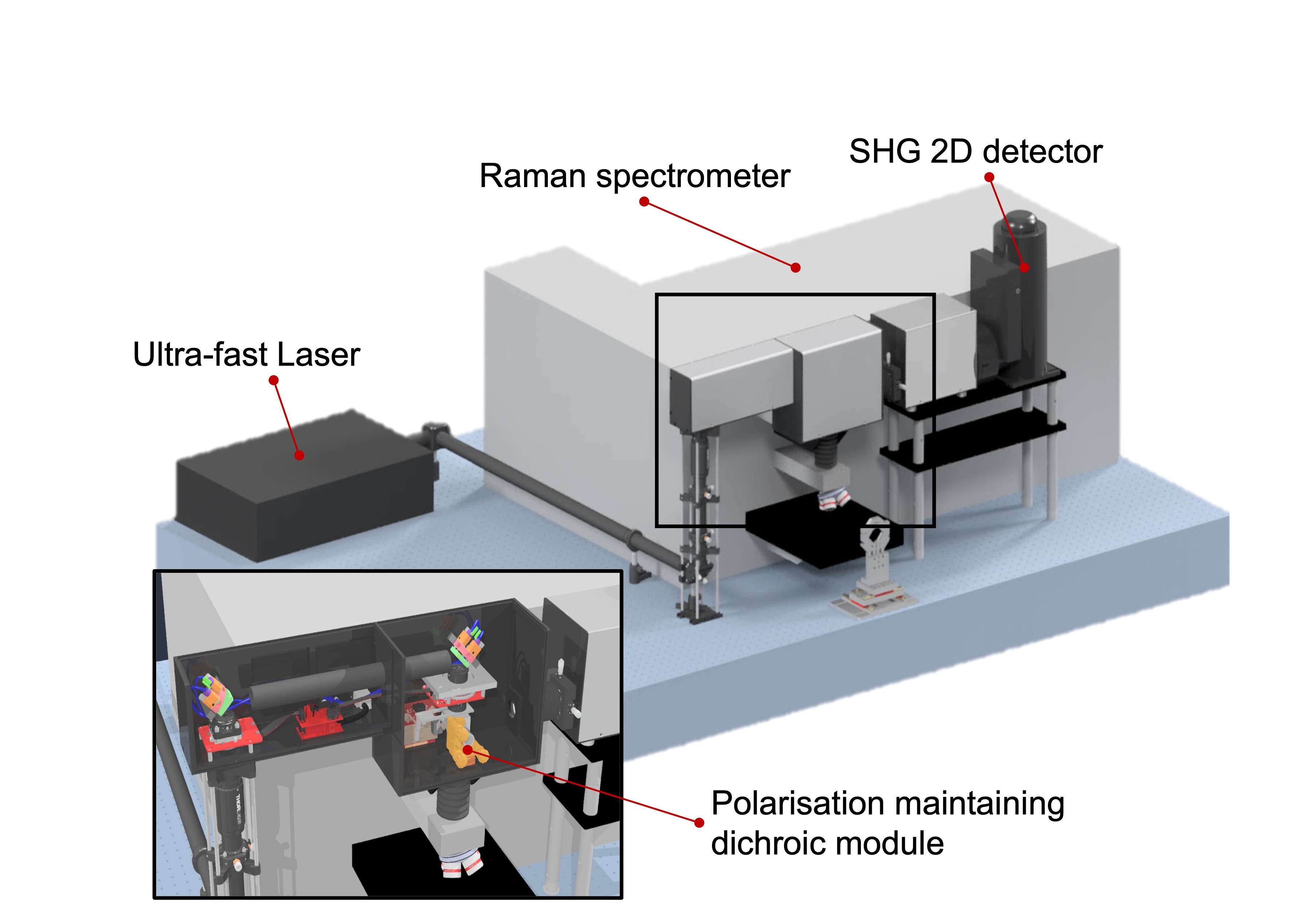}
    \caption{overview of the experimental setup for simultaneous Raman and SHG measurements with a close-up on the lasers coupling region. A detailed Scheme is provided in Figure~SM-1}
    \label{fig:setup}
\end{figure}

We developed an integrated optical platform for simultaneous polarization-resolved micro-Raman spectroscopy and Microscopic Rotational Anisotropy Second Harmonic Generation (MRA-SHG) (Fig.~\ref{fig:setup}). The system combines a commercial confocal Raman microscope with a femtosecond Ti:Sapphire–based SHG subsystem through a custom-designed polarization-maintaining dichroic beam splitter module \cite{multian_beam_2023}. The developed microscope provides precise control of excitation power and polarization, temperature-dependent measurements from 4–350 K, Raman-based thermometry with controlled laser heating, and polarization-resolved SHG detection in epi-geometry, while compensating polarization distortions along the optical path. SHG 2D patterns are collected in a dedicated detection channel using a liquid nitrogen cooled CCD (Princeton Instruments LN/400B) located in the Fourier plane of the objective. A detailed description of each subsystem and of the polarization-compensation strategy is provided in the supporting information (SI).

\subsubsection{Data Analysis}\label{subsec:data_analysis}

The 2D image collected at the Fourier plane reflects many interactions along the excitation and emission paths, involving both linear and nonlinear processes. Because a full analytical description would be impractically complex, we use a ray-tracing-based model to simulate the influence of each optical element, with details provided in SI. This approach can be summarized to few key elements:
\begin{itemize}
    \item introducing a grid of "polarized beamlet" instances characterized by positions in 3D space $\vec{r}$, complex beam parameters $q$ defining the local curvatures of the wave front, polarization states described by electrical field in form of 3D Jones vectors $\vec{E}^{\omega}$ defined in global coordinate system with corresponding wave vectors $\hat{k}^{\omega}$;
    \item propagation $\vec r$" and $\hat k$ of "polarized beamlets" in 3D space according to ray tracing rules \cite{chipman2018polarized} and transformation of electrical field $\vec E^{\omega}$ at each interface with the use of $P$-matrix approach;
    \item calculation of effective vector field distribution $\vec{E}^{\omega, \text{eff}}$ in the sample; 
    \item calculation of response electrical field $\vec{E}^{NLO}$ from effective field $\vec{E}^{\omega, \text{eff}}$;
    \item rejection of electrical field of generated $\vec{E}^{NLO}$ on the emission directions $\hat{k}^{NLO}$;
    \item propagation of emitted "polarized beamlets" through the detection channel to the detector with the analyzer oriented in parallel or crossed configuration;
    \item calculation of the Fourier images by integration of resampled beamlets intensities at the detection plane.
\end{itemize}
\begin{figure}[!ht]
    \centering
\includegraphics[width=0.95\linewidth]{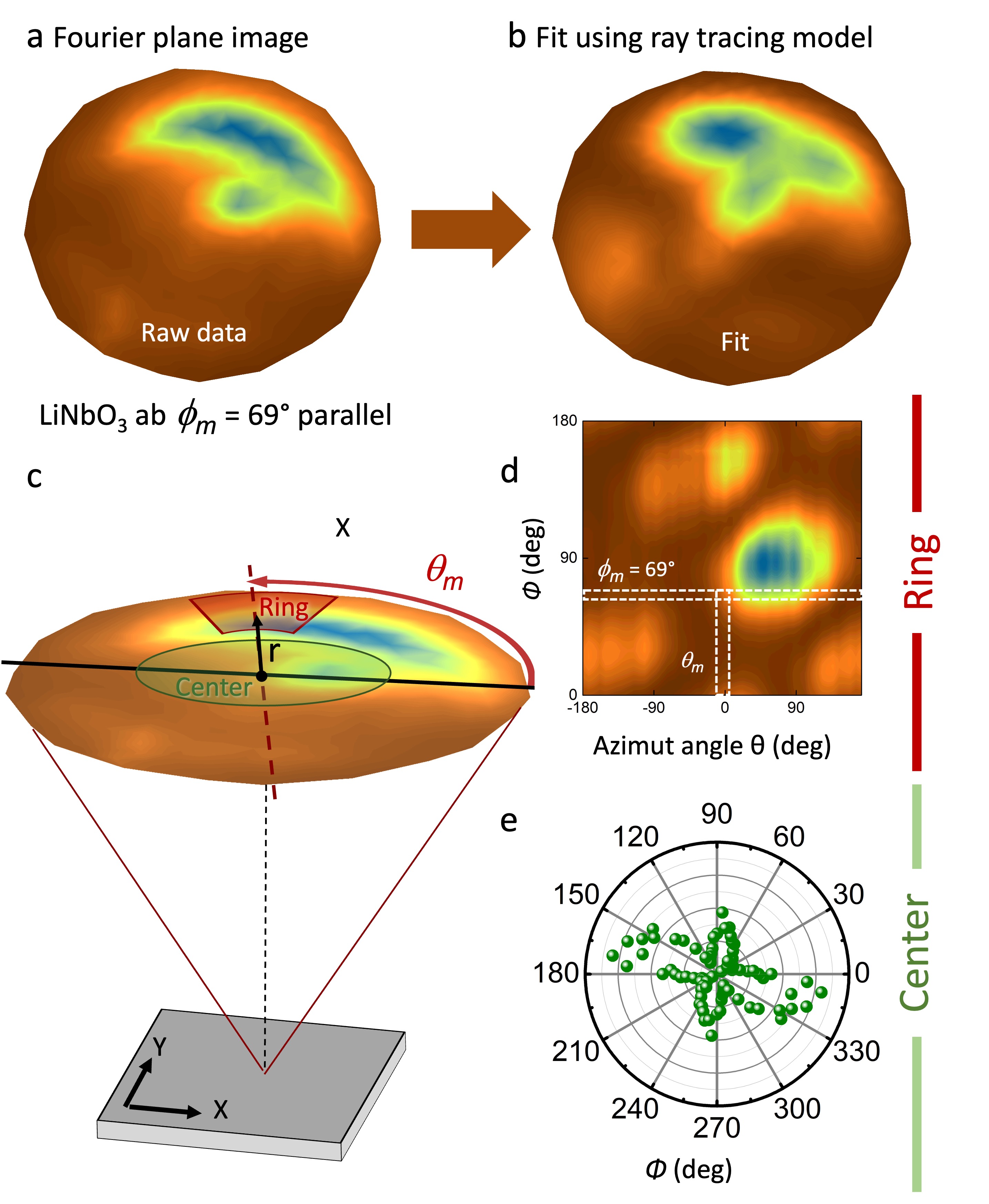}
    \caption{a) SHG image collected at Fourier plane on \ce{LiNbO3} for an incoming polarisation $\phi_n=69^o$ in parallel mode. b) fit of the Fourier image. c-d) Procedure to display the set of raw and fitted Fourier maps into a single $\phi$ / $\theta$ polar map. c) Scheme of the analyzed ring identifying angle $\phi_n$ as the angle of the incoming polarization relative to $x$ axis and $\theta_m$ the angle between the incoming polarization and the $m^{th}$ sector. d) resulting polar map after integration of all sectors relative to the two angles. e) Integrated values of the central intensity as a function of $\phi$.}
    \label{fig:data-proc}
\end{figure}

The next step is to fit parameters describing the material’s symmetry, orientation, and relevant optical properties in the ray-tracing model so that the simulated Fourier-plane patterns match the experimental ones. (Figure~\ref{fig:data-proc}a,b). Because the parameter space is high-dimensional and complex, we use a global optimization approach based on GPU-accelerated evolutionary algorithms. This allows us to fit multiple polarization configurations simultaneously and to incorporate datasets taken at different sample orientations, keeping the material properties identical, improving the robustness and accuracy of the extracted parameters (SI, Section~2).

To simplify the visualization of both raw and fitted data without inspecting each individual Fourier image, we construct a single composite image that gathers all relevant information. The procedure to go from the Fourier plane raw and fitted data to these new polar maps is detailed in Figure~\ref{fig:data-proc}c-e. For an incident light with a polarization angle $\phi_n$, we analyze each Fourier image and extract intensity by integration of a region at fixed radius and azimuth angles $\theta_m$ to form the map at the coordinates ($r$,$\theta_m$). The integration of signal intensity of the central part of the images in the Fourier plane as a function of incident polarisation defines a polar plot that represents in-plane response (see Figure~\ref{fig:data-proc}e).

Intensity $I_{r}$, $I_{c}$  integrated in "ring" and "center" regions of Fourier image for parallel or crossed configuration can be written as:
\begin{align} 
    I_{r}(r_r, \theta_m, \phi_m) &= \int_{r_r-\Delta r}^{r_r+\Delta r}\int_{\theta_m-\Delta \theta}^{\theta_m+\Delta \theta}I(r, \theta, \phi_n)d\theta dr, \\
    I_{c}(r_c, \phi_n) &= \int_{0}^{r_c}\int_{0}^{2\pi}I(r, \theta, \phi)d\theta dr, 
\end{align}
where $(r_r, \theta_m)$ - position of the integrated sector of the "ring" with size $2\Delta r \times 2\Delta \theta$, $r_c$ - radius of the central region.

\subsection{Experimental results}\label{sec:KDP}

We first acquired Raman and SHG data from KH$_2$PO$_4$ (KDP), a well-characterized nonlinear crystal with a known $\chi^{(2)}$ tensor \cite{KDP_properties}. Its low-temperature phase transition was used to correlate the temperature dependence of the Raman response with that of second-harmonic generation. Polarimetric SHG measurements were then employed to calibrate the system, enabling quantitative extraction of the nonlinear tensor elements.

The approach was subsequently applied to \ce{LiNbO3} (Z-cut, $10\times10\times0.5$mm), to demonstrate the robustness of RA-SHG for determining absolute tensor element values.

\subsubsection{Raman-SHG correlation}

At room temperature, KH$_2$PO$_4$ (KDP) is in a para-electric phase (P-KDP) with space group $I\overline{4}2d$ ($D_{2d}$) and ferroelectric phase (F-KDP) below $T_c = 123$ K with space group $Fdd2$ ($C_{2v}$). KDP is known for high transmittance, high damage threshold, and is available in large single crystal dimensions, but it has low SHG efficiency due to relatively weak nonlinear coefficients at room temperature ($d_{36} \approx 0.4$ pm/V @1064 nm \cite{KDP_properties}). 

\begin{figure}[ht!]
    \centering
    \includegraphics[width=1\linewidth]{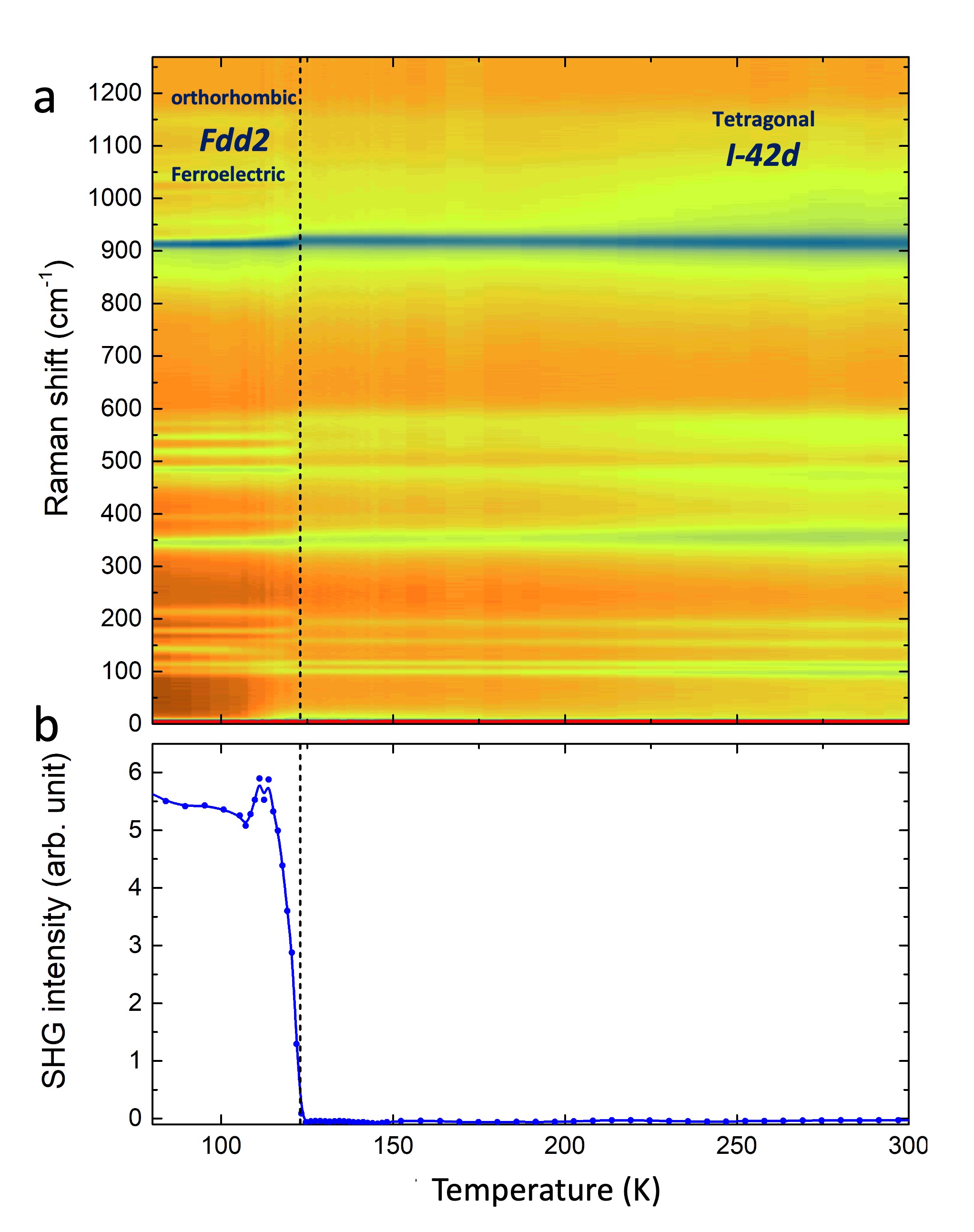}
    \caption{a) Raman and b) global SHG temperature dependence of a KDP crystal. The dashed line corresponds to the temperature of the structural transition.}
    \label{fig:exp-raman}
\end{figure}

The KDP crystals used in this study were grown from saturated water solutions and annealed to remove internal moisture \cite{kolybayeva_increase_1995}. Figure~\ref{fig:exp-raman} shows Raman (a) and total SHG (b) temperature-dependent signal recorded simultaneously from the same sample spot. Temperatures extracted from the Stokes/anti-Stokes fit of Raman spectra recorded when both excitation lasers are on match those from the cryostat sensor, confirming negligible laser heating due to low crystal absorption. As temperature elevation due to light absorption in the material depends on many parameters like absorbance, thermal conductivity, or thermal coupling with the substrate, using the Raman signal is a reliable way to directly estimate the focal point's temperature.

By cooling down below the phase transition temperature ($T = 123$ K, dashed line in Figure~\ref{fig:exp-raman}a,b), SHG increases by two orders of magnitude as expected when ferroelectric polarization sets-in. Concomitantly, Raman spectra reveal a clear lowering of symmetry through an increase in the number of active Raman modes. 

To the best of our knowledge, this is the first time that Raman spectroscopy has been conclusively used as a thermometer to verify the temperature of a distinct optical process.

\subsubsection{Multiple entry calibration}

MRA-SHG was performed at room temperature on prismatic sectors of the crystal of the tetragonal phase ($\approx 3 \times5\times$ 2 mm) of KDP for different orientations. They show a bright intensity "ring" and a faint central spot, which, although symmetry-forbidden for this material, can be attributed to slight misalignment in the normal-incidence setup. 

The proposed data analysis allows to jointly analyze the responses from the studied material and a calibration sample, measured under identical conditions. In addition, different measurements such as SHG and Raman can be combined because they are acquired from the same sample position. This shared geometry constrains the sample orientation and significantly improves the robustness of tensor element determination for both SHG and Raman, even when the model involves a large number of parameters.

For the calibration of the SHG and Raman response of the instrument, we used KDP as a reference. Additionally, we fix the known geometrical parameters of the experimental setup, including distances between optical elements, parfocal and working distances of the objective, numerical aperture and entrance pupil diameter, precise retardance of the HWP, and extinction ratio of the attenuator. The remaining parameters, like tensor elements, Jones matrix elements of objectives, orientation of the crystals, etc., are globally fitted.

\begin{figure*}[ht!]
    \centering
\includegraphics[width=0.75\textwidth]{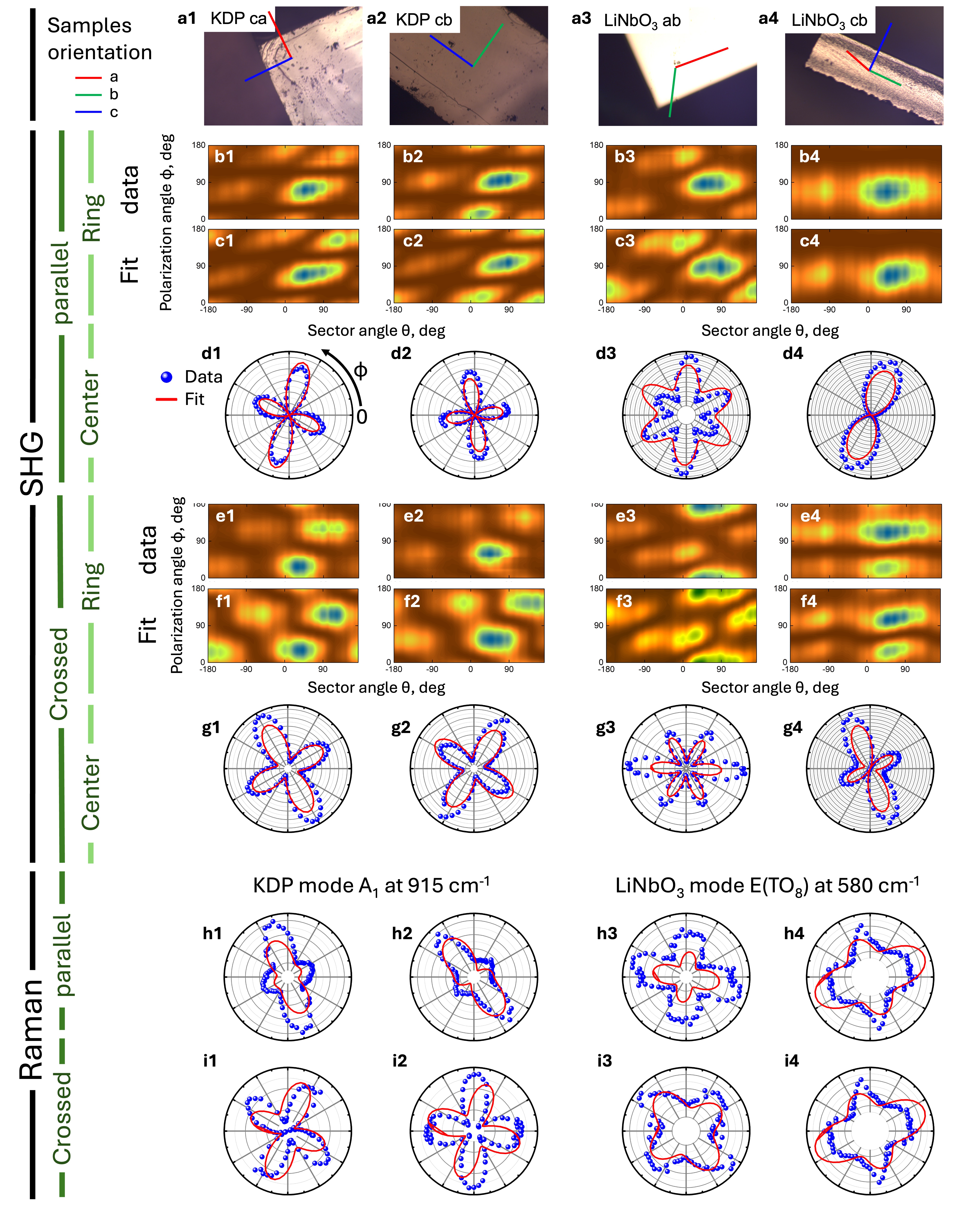}
        \caption{Full set of polarization resolved SHG (\textbf{b-g}) and Raman (\textbf{h,i}) data with the corresponding fit using the unified model described in the text. Each column corresponds to KDP (\textbf{1,2}) or \ce{LiNbO3} (\textbf{3,4}) crystals with specific orientation. \textbf{a}) Optical images of the crystals with crystallographic axes (a red, b green, c blue). \textbf{b} and \textbf{e}, experimental SHG intensity maps in parallel and crossed configurations, respectively, extracted from the "ring" part of images in the Fourier plane. \textbf{c} and \textbf{f}, corresponding fits. \textbf{d} and \textbf{g}, polarization-resolved SHG intensities in parallel and crossed configurations, respectively, extracted from the "central" part of the images in the Fourier plane. \textbf{h} and \textbf{i}, polarization-resolved Raman scattering intensities measured in parallel and crossed configurations, respectively. (\textbf{h,i;1,2}), A$_1$ Raman mode of KDP at 915 cm$^{-1}$ and (\textbf{h,i;3,4}), E Raman mode of \ce{LiNbO3} at 580 cm$^{-1}$ for corresponding orientation of the crystals. Legends and axes of all polar plots match those in d1.
} \label{fig:KDP_LiNbO3_fit}
\end{figure*}

Figure~\ref{fig:KDP_LiNbO3_fit} presents polarization maps and plots extracted from both raw and fitted SHG Fourier-plane patterns, along with polar plots of Raman intensity acquired from reference KDP and \ce{LiNbO3} crystals, both with two distinct orientations.

Quality of the orientation determination can be concluded from the agreement of fitted crystal axes orientation with those on the optical image presented in the first row of Figure~\ref{fig:KDP_LiNbO3_fit}.

The datasets for a given material were fitted with a unique nonlinear optical tensor, transformed by a tensor rotation procedure (SI) to account for the different crystal orientations. Small misalignment of the surface orientation is taken into account by considering a nonzero angle between the optical axis and the normal to the surface (SI). While each individual data input captures only a fraction of the complete response, combining multiple inputs (such as different polarization states and orientations) enables a precise determination of tensor elements, including those that are typically inaccessible in conventional measurement geometries.

\ce{LiNbO3} belongs to the $C_{3v}$ point group with $\chi^{(2)}$ tensor:
\begin{equation}
    \chi^{(2)}(C_{3v}) = \left(
    \begin{array}{ccc}
        \left[
            \begin{array}{ccc}
                 0& xyx& yzy\\
                xyx& 0& 0\\
                yzy& 0& 0
            \end{array}
        \right],\\
        \left[
            \begin{array}{ccc}
                xyx& 0& 0\\
                0& -xyx& yzy\\
                0& yzy& 0
            \end{array}
        \right],\\
        \left[
            \begin{array}{ccc}
                zyy& 0& 0\\
                0& zyy& 0\\
                0& 0& zzz
            \end{array}
        \right]
    \end{array}
    \right) \label{eq:C_3v_chi2}
\end{equation}

Assuming Kleinman symmetry, the number of free parameters in the tensor can be reduced to $3$. In our case, this assumption holds because we excite the system with $1.55$~$eV$ ($800$~$nm$) photons, well below \ce{LiNbO3} indirect bandgap ($3.3$ $eV$ ), even when two-photon absorption is considered.\cite{nikogosian_nonlinear_2005} In this case, the ($zyy$) and ($yzy$) elements become identical. 

Table \ref{tab:LiNBO3_chi2} presents a comparison between tensor elements reported in the literature and those measured in the present study. Signs of elements are reported as published, but only relative sign differences in the tensor are relevant. 

\begin{table}[ht]
    \centering
    \begin{tabular}{c|c|c|c|c}
        \makecell{$\chi^{(2)}$\\ element }& \makecell{Wavelength \\$\lambda$, nm } & \makecell{Ref. value,\\ pm/V} & \makecell{Ref.} & \makecell{This work,\\ pm/V}\\
        \hline
        \hline
        \multirow{3}*{$zzz$} & 852 & 51.4 & \cite{shoji_absolute_1997} & \multirow{3}*{-51.5$\pm$ 2.0 }\\
            &1064 & -54.0 & \cite{roberts_simplified_1992} & \\ 
            &1058 & -83.4 & \cite{boyd_linbo3:_1964}  & \\\hline
        \multirow{3}*{$zyy$} & 852 & 9.6 & \cite{shoji_absolute_1997}& \multirow{3}*{-9.1$\pm$1.0}\\
            &1064 & -8.6 & \cite{roberts_simplified_1992}\\
            &1058 & -9.3 & \cite{boyd_linbo3:_1964}& \\\hline
         \multirow{1}*{$yzy$} &  &  & & \multirow{1}*{-9.6$\pm$1.0}\\\hline
        \multirow{2}*{$xyx$} & 1064 & -4.2 & \cite{roberts_simplified_1992} & \multirow{2}*{-1.5$\pm$ 1.0}\\
        & 1058 & -4.9 & \cite{boyd_linbo3:_1964}& \\
    \end{tabular}
    \caption{Comparison of reference data and estimated $\chi^{(2)}$ tensor elements for \ce{LiNbO3} measured under $fs$ laser excitation at 800 nm}
    \label{tab:LiNBO3_chi2}
\end{table}

Within the measurement uncertainty estimated from the fitting procedure, our results are in good agreement with previous reports. Moreover, the ($zyy$) and ($yzy$) elements exhibit identical values within the experimental error, confirming the validity of Kleinman symmetry.

In this model, SHG provides rich features that fix the crystal orientation, enabling reliable analysis of the more limited Raman response.

In theory, ($A_1$) mode of KDP under study is expected to be polarization independent, according to the selection rules for in-plane excitation of the investigated crystal orientations. It is defined by the following Raman tensor:
\begin{equation}
    R(A_1) = \left(
    \begin{array}{ccc}
         a& 0&0 \\
         0& a&0 \\
         0& 0&b \\
    \end{array}
    \right),
\end{equation}
where $a=1$ and $b=0.8$ according to \cite{huang_modeling_2023}.

Reported experimental values for LiNbO$_3$ can exhibit significant variations depending on crystal quality, fabrication method, and measurement conditions. Such variations are also observed in this work when comparing home-made microcrystals and commercial \ce{LiNbO3} samples, and are not fully captured by numerical databases often considering the highest symmetry for a given structure. Several studies have reported the presence of measurement artifacts in Raman responses \cite{huang_modeling_2023,demos_measurement_2011}, consistent with our observations. While polarization in the SHG subsystem is well preserved by the non-polarizing module, the Raman response is strongly modulated by the spectrometer. The observed polarization-dependent signal, therefore, arises from a combination of the system response and out-of-plane leakage under tight focusing.

By constraining the crystal orientation through the SHG fitting, we calibrate the polarization response of the Raman setup and subsequently extract the Raman tensor of the (E) mode in \ce{LiNbO3}. According to the computational Raman database \cite{bagheri_high-throughput_2023}, the (E) mode observed experimentally at 580~cm$^{-1}$ corresponds to the calculated one at 569~cm$^{-1}$, which is characterized by the following Raman tensor:

\begin{equation}
    R(E@569cm^{-1}) = \left(
    \begin{array}{ccc}
    -1.5&       1.3&      -3.4\\
     1.3&       1.5&       3.8\\
    -3.4&       3.8&       0.0\\
    \end{array}
    \right),
\end{equation}
We perform the fitting with an arbitrary symmetric Raman tensor with 6 free parameters:
\begin{equation}
    R_{sym} = \left(
    \begin{array}{ccc}
        a& d& e\\
		d& b& f\\
		e& f& c\\
    \end{array}
    \right)
\end{equation}
A separate fitting exclusively applied to the Raman response with this number of free parameters and in the presence of strong distortion of polarization by the optical system would not converge to a reliable set of values. Constraining Raman by the input from SHG by our comprehensive approach significantly improves the result.

Fitting gives us the following results, normalized to match the $a$ element of the calculated tensor:
\begin{equation}
    R(E @ 580cm^{-1})_{fit} = \left(
    \begin{array}{ccc}
    -1.5 &  1.8& -0.3\\
     1.8 &  1.8&  2.5\\
    -0.3&  2.5& -2.9\\
    \end{array}
    \right).
\end{equation}
Most of the fitted tensor elements are in good agreement with the calculated values. The largest discrepancies are observed for the $c$ and $e$ elements, which correspond to directions along the optical axis of the crystal. In addition, the fitting procedure correctly reproduces the sign of most tensor elements. The differences in the $c$, $e$, and $f$ elements may be attributed to variations in surface treatment among the investigated \ce{LiNbO3} crystal faces, which can lead to changes in Raman peak amplitudes, as reported in Ref.\citenum{galinetto_micro-raman_2007}.

\section{Conclusion}

In this study, we developed an experimental setup to measure Raman spectroscopy and second harmonic generation simultaneously, and rapidly and quantitatively extract the second-order nonlinear optical tensor elements of a material. This technique takes advantage of the spatial and temporal mismatch of a Bessel-Gauss beam at the laser focal point to simultaneously measure SHG at normal and fixed incidence. Thanks to the development of a patented dichroic module that preserves the polarization state, the method was implemented on a confocal microscope and enables full polarization dependence measurements on microscopic samples, while minimizing movements of both the sample and optical components. 

A numerical data analysis method, based on a ray-tracing model that accounts for each optical component of the instrument as well as the optical parameters of the sample, enables the full use of the light intensity distribution collected in the Fourier plane of the objective.

We fitted the SHG and Raman responses of \ce{LiNbO3} together with the corresponding responses of KDP reference crystals and successfully extracted the nonlinear tensor elements. The values obtained are in close agreement with those reported in the literature\cite{KDP_properties, shoji_absolute_1997, roberts_simplified_1992, boyd_linbo3:_1964}.

This new method offers a powerful and precise way to measure the second-order nonlinear optical susceptibility tensor for SHG, with the possibility of coupling with other responses like Raman scattering. With further development of field propagation inside the samples with different topologies, it can potentially be applied to higher-order nonlinear processes and materials with complex structures, like a stack of films, nano-objects, and metamaterials. One of the promising applications is the rapid localization and characterization of 2D materials, which could greatly accelerate the fabrication of complex heterostructures. Inspired by X-ray diffraction or the structural biology Cryo-EM’s approach \cite{Dubochet1988CryoEM}, the technique could be used for global tensor fitting of randomly oriented dielectric nanoparticles for rapid characterization.


%
%

%

\section*{Data Availability Statement}
Data that support the findings of this study are openly available in Yareta at: \\ \href{https://doi.org/10.26037/yareta:esl6qjw4yva6towy2hch7o2h3q}{https://doi.org/10.26037/yareta:esl6qjw4yva6towy2hch7o2h3q}

\section*{Supplementary Information}
Supplementary.pdf file contains a detailed description of the experimental measurement setup, as well as a step-by-step description of the construction of the ray-tracing model.
\begin{acknowledgments}

This project was supported by the Swiss National Science Foundation through projects CRSK-2-227400

\end{acknowledgments}
\bibliography{biblio}

\end{document}